\documentclass{elsart}
\usepackage{amsmath}
\usepackage{amssymb}
\usepackage{epsfig}
\usepackage{url}

\newcommand{\aff}[2]{Dipartimento di Fisica dell'Universit\`a #1 e Sezione INFN, #2, Italy.}
\newcommand{\affd}[1]{Dipartimento di Fisica dell'Universit\`a e Sezione INFN, #1, Italy.}

\newcommand{\DAFNE}{DA\char8NE}
\newcommand{\SN}[2]{\ensuremath{#1\times10^{#2}}}
\newcommand{\bra}[1]{\ensuremath{\left|#1\right>}}
\renewcommand{\to}{\ensuremath{\rightarrow}}

\begin{document}
\begin{frontmatter}

\title{Measurement of the pseudoscalar mixing angle and $\boldsymbol{\eta^{\prime}}$ gluonium content with the KLOE detector}

\collab{The KLOE Collaboration}

\author[Na]{F.~Ambrosino},
\author[Frascati]{A.~Antonelli},
\author[Frascati]{M.~Antonelli},
\author[Roma3]{C.~Bacci},
\author[Karlsruhe]{P.~Beltrame},
\author[Frascati]{G.~Bencivenni},
\author[Frascati]{S.~Bertolucci},
\author[Roma1]{C.~Bini},
\author[Frascati]{C.~Bloise},
\author[Roma3]{S.~Bocchetta},
\author[Roma1]{V.~Bocci},
\author[Frascati]{F.~Bossi},
\author[Frascati,Virginia]{D.~Bowring},
\author[Roma3]{P.~Branchini},
\author[Roma1]{R.~Caloi},
\author[Frascati]{P.~Campana},
\author[Frascati]{G.~Capon},
\author[Na]{T.~Capussela},
\author[Roma3]{F.~Ceradini},
\author[Frascati]{S.~Chi},
\author[Na]{G.~Chiefari},
\author[Frascati]{P.~Ciambrone},
\author[Virginia]{S.~Conetti},
\author[Frascati]{E.~De~Lucia},
\author[Roma1]{A.~De~Santis},
\author[Frascati]{P.~De~Simone},
\author[Roma1]{G.~De~Zorzi},
\author[Frascati]{S.~Dell'Agnello},
\author[Karlsruhe]{A.~Denig},
\author[Roma1]{A.~Di~Domenico},
\author[Na]{C.~Di~Donato\thanksref{*}},
\author[Pisa]{S.~Di~Falco},
\author[Roma3]{B.~Di~Micco},
\author[Na]{A.~Doria},
\author[Frascati]{M.~Dreucci},
\author[Frascati]{G.~Felici},
\author[Frascati]{A.~Ferrari},
\author[Frascati]{M.~L.~Ferrer},
\author[Frascati]{G.~Finocchiaro},
\author[Roma1]{S.~Fiore},
\author[Frascati]{C.~Forti},
\author[Roma1]{P.~Franzini},
\author[Frascati]{C.~Gatti},
\author[Roma1]{P.~Gauzzi},
\author[Frascati]{S.~Giovannella},
\author[Lecce]{E.~Gorini},
\author[Roma3]{E.~Graziani},
\author[Pisa]{M.~Incagli},
\author[Karlsruhe]{W.~Kluge},
\author[Moscow]{V.~Kulikov},
\author[Roma1]{F.~Lacava},
\author[Frascati]{G.~Lanfranchi},
\author[Frascati,StonyBrook]{J.~Lee-Franzini},
\author[Karlsruhe]{D.~Leone},
\author[Frascati]{M.~Martini},
\author[Na]{P.~Massarotti},
\author[Frascati]{W.~Mei},
\author[Na]{S.~Meola},
\author[Frascati]{S.~Miscetti},
\author[Frascati]{M.~Moulson},
\author[Frascati]{S.~M\"uller},
\author[Frascati]{F.~Murtas},
\author[Na]{M.~Napolitano},
\author[Roma3]{F.~Nguyen},
\author[Frascati]{M.~Palutan},
\author[Roma1]{E.~Pasqualucci},
\author[Roma3]{A.~Passeri},
\author[Frascati,Energ]{V.~Patera},
\author[Na]{F.~Perfetto},
\author[Roma1]{L.~Pontecorvo},
\author[Lecce]{M.~Primavera},
\author[Frascati]{P.~Santangelo},
\author[Roma2]{E.~Santovetti},
\author[Na]{G.~Saracino},
\author[Frascati]{B.~Sciascia},
\author[Frascati,Energ]{A.~Sciubba},
\author[Pisa]{F.~Scuri},
\author[Frascati]{I.~Sfiligoi},
\author[Frascati]{T.~Spadaro},
\author[Roma1]{M.~Testa},
\author[Roma3]{L.~Tortora},
\author[Roma1]{P.~Valente},
\author[Karlsruhe]{B.~Valeriani},
\author[Frascati]{G.~Venanzoni},
\author[Roma1]{S.~Veneziano},
\author[Lecce]{A.~Ventura},
\author[Frascati]{R.Versaci},
\author[Frascati,Beijing]{G.~Xu}

\address[Frascati]{Laboratori Nazionali di Frascati dell'INFN, 
Frascati, Italy.}
\address[Karlsruhe]{Institut f\"ur Experimentelle Kernphysik, 
Universit\"at Karlsruhe, Germany.}
\address[Lecce]{\affd{Lecce}}
\address[Na]{Dipartimento di Scienze Fisiche dell'Universit\`a 
``Federico II'' e Sezione INFN,
Napoli, Italy}
\address[Pisa]{\affd{Pisa}}
\address[Energ]{Dipartimento di Energetica dell'Universit\`a 
``La Sapienza'', Roma, Italy.}
\address[Roma1]{\aff{``La Sapienza''}{Roma}}
\address[Roma2]{\aff{``Tor Vergata''}{Roma}}
\address[Roma3]{\aff{``Roma Tre''}{Roma}}
\address[StonyBrook]{Physics Department, State University of New 
York at Stony Brook, USA.}
\address[Virginia]{Physics Department, University of Virginia, USA.}
\address[Beijing]{Permanent address: Institute of High Energy 
Physics of Academica Sinica,  Beijing, China.}
\address[Moscow]{Permanent address: Institute for Theoretical 
and Experimental Physics, Moscow, Russia.}
\thanks[*]{Corresponding author.\\
           {\it E-mail address:} camilla.didonato@na.infn.it (C. Di Donato).}

\begin{abstract}
We have measured the ratio $R_{\phi}=\text {BR}(\phi \to \eta^{\prime} \gamma)/
\text {BR}(\phi \to \eta \gamma)$ by looking for the radiative decays
$\phi \to \eta^{\prime} \gamma$ and $\phi \to \eta \gamma$ into the 
final states $\pi^+\pi^-7\gamma$ and $7\gamma$, respectively, in a sample of
$\sim$\SN{1.4}{9} $\phi$ mesons produced at the Frascati $\phi$ factory.
We obtain $R_{\phi}=\SN{(4.77\pm0.09_\text{stat}\pm0.19_\text{syst})}{-3}$, from which we derive
${\rm BR}(\phi \to \eta^{\prime} \gamma)=\SN{(6.20\pm0.11_\text{stat}\pm0.25_\text{syst})}{-5}$.
Assuming the $\eta^\prime$ has zero gluonium content, we extract the pseudoscalar mixing angle
in the quark-flavor basis, $\varphi_P=(41.4\pm0.3_\text{stat}\pm0.7_\text{syst}\pm0.6_\text{th})^{\circ}$.
Combining the value of $R_{\phi}$ with other constraints, we estimate
the fractional gluonium content of the $\eta^{\prime}$ to be $Z^2 = 0.14\pm0.04$ and the mixing angle to be
$\varphi_P = (39.7\pm0.7)^{\circ}$.

\end{abstract}


\end{frontmatter}

% main text
\section{Introduction}
\label{intro}
The value of the $\eta$-$\eta^{\prime}$ mixing angle in the pseudoscalar meson nonet, $\theta_P$,
has been discussed extensively over the last 35 years, and is today one of the most
interesting SU(3)-breaking hadronic parameters to measure \cite{Ben}.
In the context of chiral perturbation theory, it has been demonstrated that a description of the
$\eta$-$\eta^{\prime}$ system beyond leading order cannot be achieved in terms of just one mixing
angle \cite{Leutw}. However, in the flavor basis the two mixing angles are equal, apart from terms
which violate the OZI-rule; thus working in this basis it is still possible to use a single mixing angle,
$\varphi_{P}$ \cite{Feld}.
The ratio $R_{\phi}$ of the two branching ratios $\phi \to \eta^{\prime} \gamma$ and
$\phi \to \eta \gamma$ can be related to the $\eta$-$\eta^{\prime}$ mixing angle in 
the flavor basis \cite{Feld,Becchi,Rosner,Ball,Bramon,Rafael} and to the gluonium content
of the $\eta^{\prime}$ meson \cite{Rosner,kou}.

In this work we present a measurement of $R_{\phi}$ obtained using an integrated luminosity
of $427$~pb$^{-1}$ collected by KLOE during the years 2001--2002.
The best measurement available to date was also published by KLOE \cite{Fabio2},
and was obtained from the analysis of the $\phi \to \eta^{\prime} \gamma$ decay with the
$\pi^+ \pi^-3\gamma$ final state and the $\phi \to \eta \gamma$ decay with the
$\pi^+ \pi^- \pi^0 \gamma$ final state. The earlier KLOE measurement was based on an integrated luminosity of about $16$ $\text{pb}^{-1}$,
collected during the year 2000. Previous measurements were performed by the SND and CMD-2
collaborations \cite{SND,CMD2}.

\section{\DAFNE\ and KLOE}
\label{kloe}
 
The Frascati $\phi$ factory \DAFNE\ is an $e^+e^-$ collider running at a center-of-mass energy
$\sqrt{s}=1.02$ GeV, where $\phi$ mesons are produced with the beams colliding at a
small crossing angle (25 mrad).
The KLOE detector consists of two main subdetectors: a large cylindrical drift chamber \cite{DC}
and a sampling lead/scintillating-fiber electromagnetic calorimeter \cite{EmC}.
A superconducting coil surrounding the calorimeter provides a solenoidal
field of 0.52 T.

The drift chamber is 3.3 m in length and has a 2 m radius; it has full stereo geometry and
operates with a gas mixture of $90\%$ helium$/10\%$ isobutane.
The momentum resolution is $\sigma(p_{\perp})/(p_{\perp})\leq 0.4\%$; the spatial resolution is
$\sigma_{xy}\simeq 150$ $\mu\text{m}$ and $\sigma_{z}\simeq 2$ $\text{mm}$.
Vertices are reconstructed with a spatial resolution of $\simeq 3$ $\text{mm}$.

The calorimeter is divided into a barrel and two endcaps; it surrounds the drift chamber and covers 
$98\%$ of the solid angle. The calorimeter is segmented into $2440$ cells of cross section $4.4\times4.4$
$\text{cm}^{2}$ in the plane perpendicular to the fibers.
Each cell is read out at both ends by photomultiplier tubes. Particle arrival times and the three-dimensional positions of the energy deposits are obtained from the signals collected at the two ends;
cells close in time and space are grouped into a calorimeter cluster.
The cluster energy $E$ is the sum of the cell energies, while the cluster time $T$ and position $R$
are energy-weighted averages. The probability for a photon to fragment into
two or more clusters (splitting) is taken into account during event reconstruction by a dedicated
procedure. The energy resolution is $\sigma_E/E=5.7\%/\sqrt{E(\text{GeV})}$; 
the timing resolution is $\sigma_t=57$ ps $/\sqrt{E(\text{GeV})}\oplus 100$ ps.
The KLOE trigger \cite{TR} is based on calorimeter and chamber information.

\section{Event selection}
\label{es}
The analysis has been performed using a sample of $\sim$\SN{1.4}{9}
$\phi$ mesons collected in 2001--2002.
We search for two categories of events. The first, 
$\phi\to \eta^{\prime}\gamma \to \pi^+\pi^- 7\gamma$, is produced 
through two different decay chains: 
\begin{gather}
\phi\rightarrow \eta^{\prime}\gamma\,\,\,\text{with}\,\,\,\eta^{\prime}\rightarrow\pi^+\pi^-\eta
\,\,\, \text{and}\,\,\,\eta\rightarrow\pi^0 \pi^0 \pi^0 \label{ch} \\
\phi\rightarrow \eta^{\prime}\gamma\,\,\,\text{with}\,\,\,\eta^{\prime}\rightarrow\pi^0\pi^0\eta
\,\,\,\text{and}\,\,\,\eta\rightarrow \pi^+\pi^-\pi^0.  \label{neu}
\end{gather}
The second, $\phi \to \eta \gamma \to 7\gamma$, is produced through
\begin{equation}
\phi\rightarrow \eta \gamma\,\,\,\text{with}\,\,\,\eta \rightarrow\pi^0\pi^0\pi^0.\label{eta}
\end{equation}
The latter channel is used for normalization of the rates and is practically background free \cite{KN}. 
Processes with $\phi \to K_SK_L$, where the $K_L$ decays near the beam interaction point (IP)
can mimic the final state with $\pi^+\pi^- 7\gamma$ because of the presence of an additional
photon due either to machine background or split clusters.
The most important decay chains of this type are the following: 
\begin{gather}
\phi\rightarrow K_S K_L\,\,\,\text{with}\,\,\,K_S \rightarrow\pi^+\pi^-\,\,\,\text{and}\,\,\,K_L \to \pi^0\pi^0\pi^0 \label{bg1} \\
\phi\rightarrow K_S K_L\,\,\,\text{with}\,\,\,K_S \rightarrow\pi^0\pi^0\,\,\,\text{and}\,\,\,K_L \to \pi^+\pi^-\pi^0 \label{bg2} \\
\phi\rightarrow K_S K_L\,\,\,\text{with}\,\,\,K_S \rightarrow\pi^+\pi^-\gamma\,\,\,\text{and}\,\,\,K_L \to \pi^0\pi^0\pi^0 \label{bg3}
\end{gather}
In particular, the last of these channels has the same final state used to identify $\phi\rightarrow \eta^{\prime}\gamma$.
Backgrounds from other sources are negligible.

All events must pass a first-level selection (FLS) \cite{offline}, consisting
of a machine background
filter and an event selection procedure that assigns events into categories.
After the FLS, we select $\phi \rightarrow \eta \gamma$ (normalization) events
by requiring the presence of seven
clusters not associated to tracks. Each cluster must have $E_{\gamma}> 10$ MeV
and polar angle $21^{\circ}<\theta_{\gamma}<159^{\circ}$ with respect to the beam direction, in order to exclude
the focusing quadrupoles.
Each cluster time must be compatible with the hypothesis of a photon coming from the IP,
so that the requirement to be satisfied is $|T_{\gamma}-R_{\gamma}/c|<
5\sigma_t$,
where $T_{\gamma}$ and $R_{\gamma}$ are the cluster time and position and $\sigma_t$ is the
time resolution. We further require that no tracks come from the interaction region.
To select $\phi\rightarrow \eta^{\prime}\gamma$ events, we similarly require seven prompt photons; in addition, we require the presence of a vertex formed 
by two tracks of opposite charge within a cylindrical region
$\sqrt{x_\text{vtx}^2 + y_\text{vtx}^2}<4~\text{cm}$ and $|z_\text{vtx}|< 8~\text{cm}$ around the IP.
After applying these selections, we perform a kinematic fit requiring energy-momentum conservation, and
times and path lengths to be consistent with the speed of light for photon candidates.
We perform an inclusive measurement of the two different decay chains contributing to the $\phi\rightarrow \eta^{\prime}\gamma$ process.

The background is significantly reduced by the event-classification cuts
in the FLS, which efficiently identify $K_S$ decays. 
By selecting events classified only as pure radiative decays,
the $K_S K_L$ backgrounds are largely reduced.
After this cut, background from processes (\ref{bg1}) and (\ref{bg3}) above is 
practically negligible, while to further 
reduce background from process (\ref{bg2}), we require 
$E_{\gamma}> 20~\text{MeV}$ for each of the seven clusters, where the 
improved estimates of the cluster energies from the kinematic fit are 
used to evaluate this cut.

To estimate the absolute number of residual background events,
the number of $\phi$ mesons in the sample has been evaluated.
We obtain $N_{\phi} = (1.37 \pm 0.02) \times 10^9$, where
the uncertainty comes from the estimate of the
integrated luminosity and the experimentally-determined value of the 
leptonic decay width of the $\phi$.
The luminosity is known with a total error of $0.6\%$ \cite{lumy},  
while the leptonic width has been measured by KLOE with a $1.7\%$ 
uncertainty \cite{Mario}.

For each background channel, Table~\ref{tabef} lists the branching ratio
multiplied by the overall selection efficiency as determined
by Monte Carlo (MC) simulation.
The branching ratio for process (\ref{bg3}) has been evaluated using
$\text {BR}(K_S \to \pi^+\pi^-\gamma)=(4.87\pm0.11)\times 10^{-3}$, 
for $E_{\gamma}> 20$ MeV$/c$ \cite{ramberg}, the same cut used for 
our analysis.

Residual background due to multiphoton final states with a photon conversion 
and production of a charged-particle vertex has been reduced to a negligible level by
a cut on the opening angle of the two tracks.
Requiring $\cos \theta_{\pi^+ \pi^-}<0.84$ \cite{KN} leads to a small 
effect on the signal efficiency and gives rise to a negligible contribution 
to the systematic error.
%%%%%%%%%%%%%%%%%%
\begin{table}[htb]
\vspace{+20 pt}
\caption{Branching ratios times selection efficiencies 
$\text{BR}(\phi\to K_SK_L) \cdot \text{BR}(K_S\to\pi \pi(\gamma))
\cdot \text{BR}(K_L\to\pi\pi\pi)\cdot
\varepsilon (\text{MC})$ and event yield for $\phi \to K_S K_L$ backgrounds}
\label{tabef}
\begin{tabular}{@{}lccc}
\hline
%$K_S \rightarrow$ & $\pi^+\pi^-$ & $ \pi^0\pi^0$ & $\pi^+\pi^-\gamma$\\   
%$K_L \to $& $\pi^0\pi^0\pi^0$ & $\pi^+\pi^-\pi^0$ & $ \pi^0\pi^0\pi^0$\\ \hline  
$K_S/K_L \rightarrow$ & $\pi^+\pi^-/\pi^0\pi^0\pi^0$ & $ \pi^0\pi^0/\pi^+\pi^-\pi^0$ &
 $\pi^+\pi^-\gamma/\pi^0\pi^0\pi^0$\\   \hline
BR$\times \varepsilon$(MC) &$(44\pm3)\cdot 10^{-9}$&$(112\pm4)\cdot 10^{-9}$&$(95\pm3)\cdot 10^{-9}$\\
Background events  & $60\pm4$ & $153\pm6$ & $130\pm5$\\
    \hline 
\end{tabular}\\[2pt]
\vspace{+20 pt}
\end{table}
%%%%%%%%%%%%%%%%%%%%%%%%%%%%%%%%%%%%%%%%%%%%%%%%%%%%%%%%%%%%%%%%%%%%%%%%%%%%%%%%%%%%%%

After all selection cuts, we find $3750$ events with an expected background 
of $N_\text{background}=343\pm43$. 
The background uncertainty comes from the uncertainties on
the number of $\phi$ mesons, the selection efficiency $(\pm15)$, and
the evaluation of the machine background level $(\pm28)$ \cite{KN}. 
We treat the background uncertainty as a systematic contribution to the error on $N_{\eta^{\prime} \gamma}$.
The final number of events from processes (\ref{ch}) and (\ref{neu}), after background
subtraction, is $N_{\eta^{\prime} \gamma} = 3407\pm61_\text{stat}\pm43_\text{syst}$.
The event-selection efficiency\footnote{We perform an inclusive measurement of two different decay chains, (\ref{ch}) and (\ref{neu}), that contribute to the
$\phi\rightarrow \eta^{\prime}\gamma$ process. The event-selection efficiency 
is obtained from the branching-ratio weighted average:
$\varepsilon_{\eta^{\prime}}=
(\varepsilon_{(1)}\text{BR}_{(1)}+\varepsilon_{(2)}\text{BR}_{(2)})/(\text{BR}_{(1)}+\text{BR}_{(2)})$}
after FLS cuts is evaluated by MC simulation to be
$\varepsilon_{\eta^{\prime}}=(23.45\pm0.16)\%$.
The sample of MC signal events is five times larger than the data sample and 
has been produced by carefully simulating the run conditions during 2001--2002 
data taking.
Corrections for small data-MC discrepancies have been estimated directly from control data samples and applied.
The FLS efficiency has been measured from data using a small subsample of minimum-bias
events reconstructed without the FLS filters; the result
is $\epsilon_{\eta^{\prime}}^{FLS}=(97\pm1)\%$. 

\begin{figure}[h]
\begin{center}
\begin{tabular}{cc}
\mbox{\epsfig{file=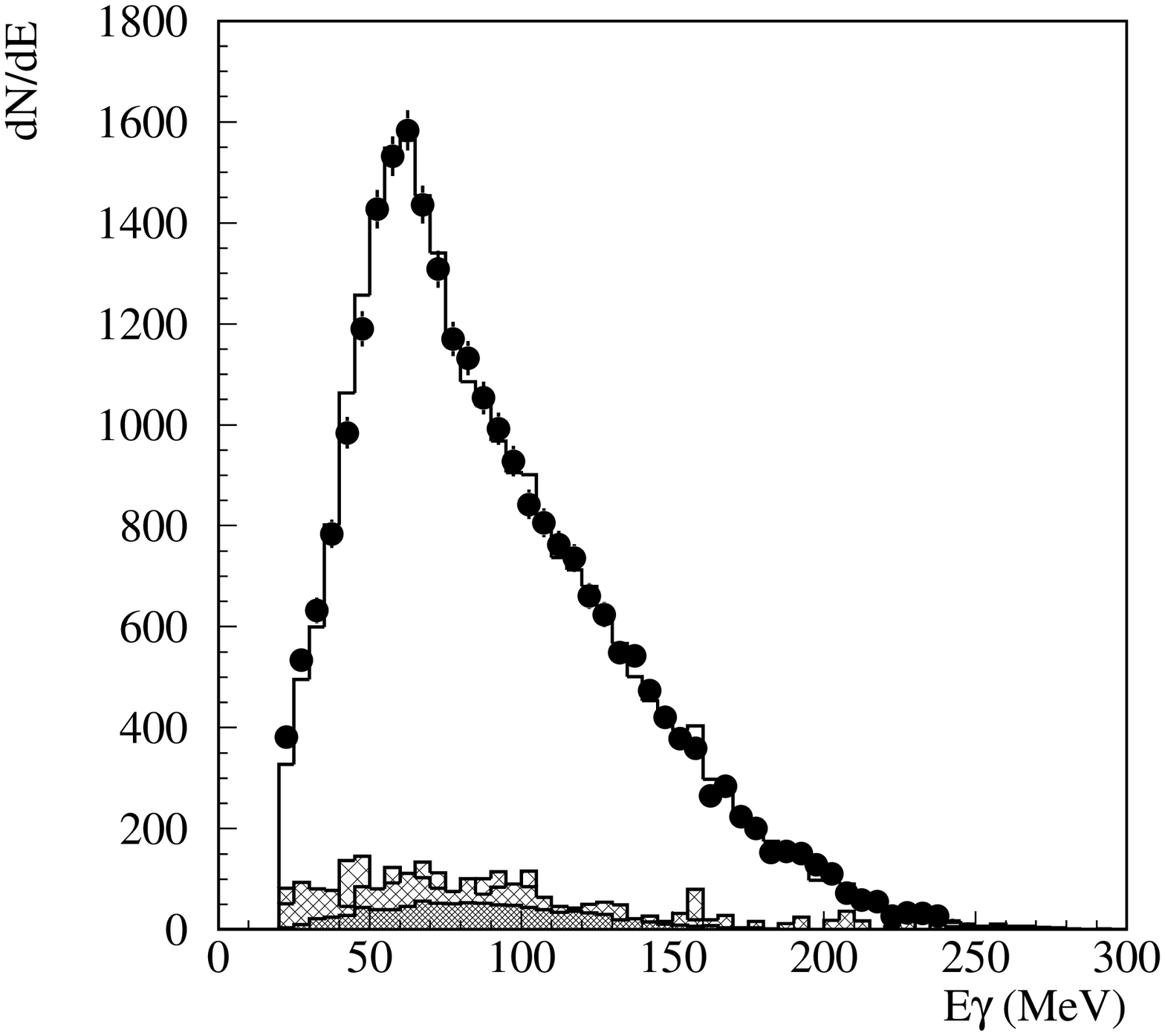, width=7.0cm}} &
\mbox{\epsfig{file=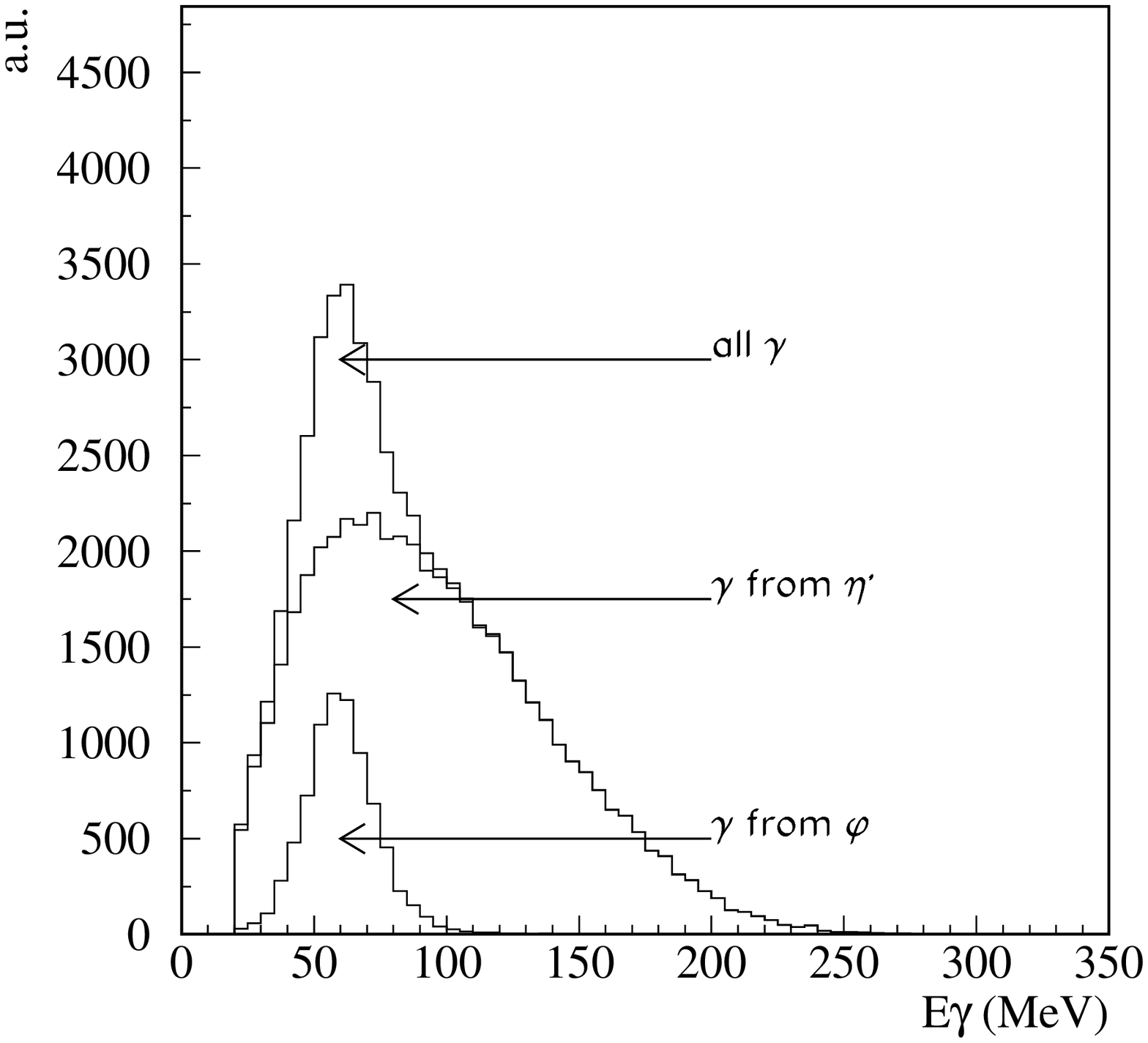, width=7.0cm}}\\
\end{tabular}
\end{center}
\caption{{\footnotesize $\phi \to \eta^{\prime} \gamma$. 
Energy distribution for all seven prompt
neutral clusters. Left: Data-MC comparison after all cuts. Dots show data; 
histograms show MC distributions for each of the three background processes, 
(\ref{bg3}), (\ref{bg2}), and (\ref{bg1}), and for the sum of signal and 
background contributions.
Right: MC $E_{\gamma}$ distributions for the recoil $\gamma$ emitted in the $\phi$ decay, the six $\gamma$'s from the $\eta^{\prime}$ decay, and the overall distribution.}}
\label{sevengamma}
\end{figure}

\begin{figure}[htb]
\begin{center} 
\begin{tabular}{cc}
\mbox{\epsfig{file=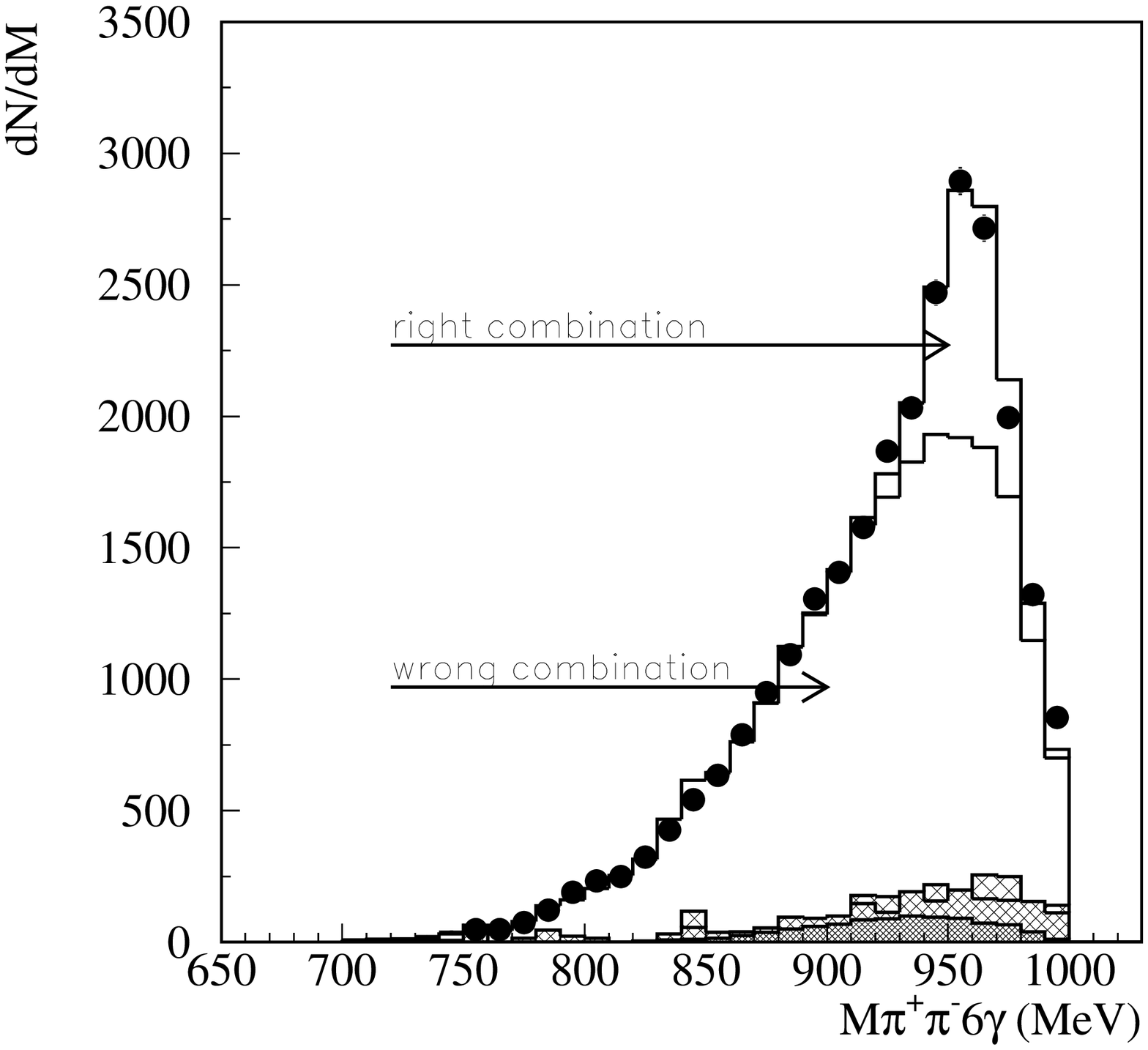,width=7.0cm}} &
\mbox{\epsfig{file=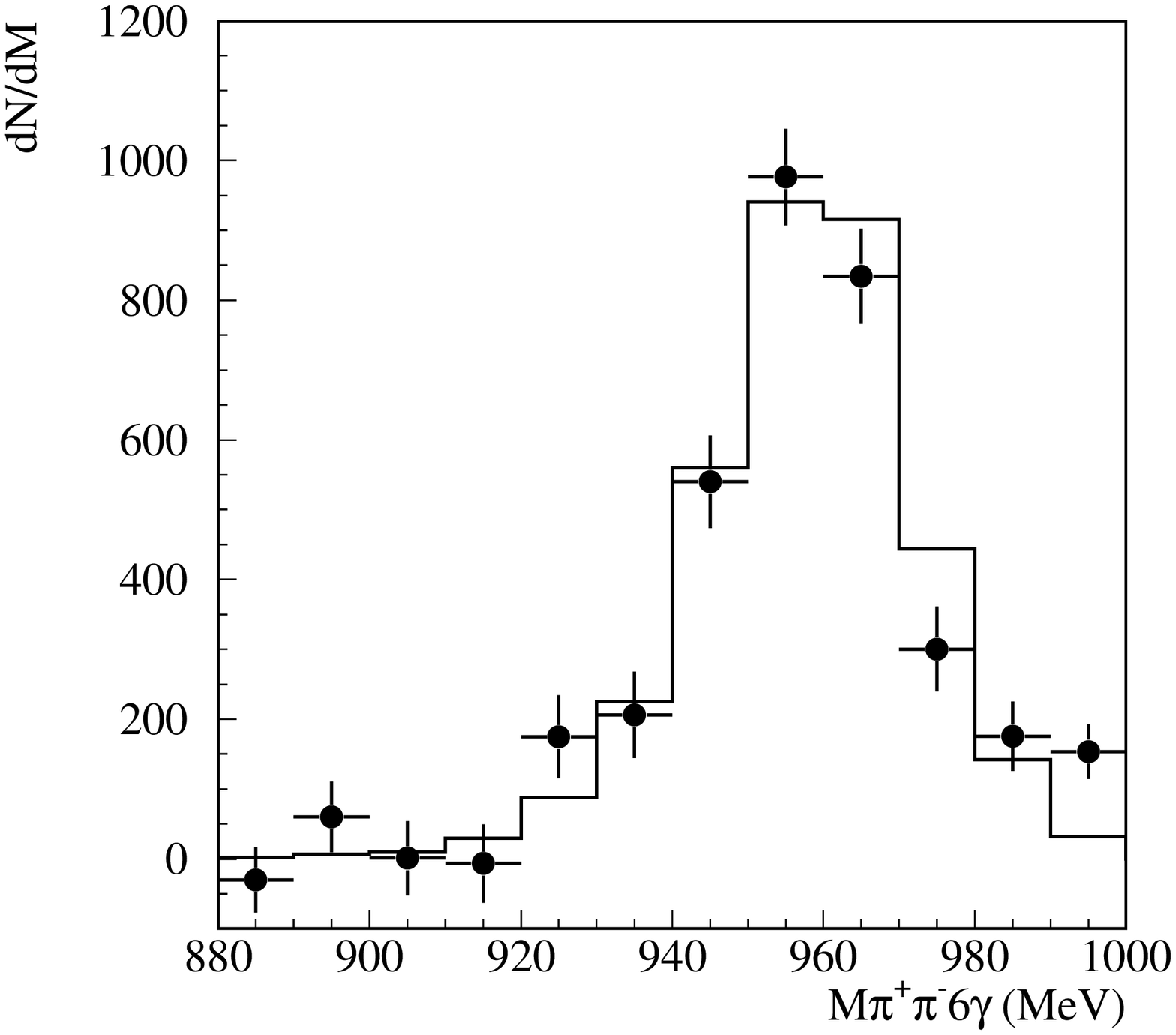,width=7.0cm}} \\
\end{tabular}
\end{center}
\caption{{\footnotesize $\phi \to \eta^{\prime} \gamma$. 
Left: Distribution of $M(\pi^+\pi^- 6\gamma)$ for all possible combinations. 
Dots are data; the histogram represents
the MC sum of the three background contributions, (\ref{bg3}), (\ref{bg2}) and 
(\ref{bg1}), the combinatorial background, and the contribution from
correctly paired signal events.
Right: Data-MC comparison of $M(\pi^+\pi^- 6\gamma)$ distributions in the
area about the peak,
after bin-by-bin subtraction.}}
\label{Minvetap}
\end{figure}

Figure~\ref{sevengamma}, left, shows the
data-MC comparison of the photon-energy spectrum after all cuts.
For $\eta^{\prime} \gamma$
events, the agreement is good.
However, it is not possible to identify the recoil photon since its spectrum overlaps with
the spectrum for photons from  $\eta^{\prime}$ decay (see Fig.~\ref{sevengamma}, right).
Therefore, when building the $\pi^+\pi^-6\gamma$ invariant mass, we have a
combinatorial background due to incorrect photon association. 
If we subtract the MC
estimate of the physical and combinatorial background from data,
we obtain a peak at the expected value of the $\eta^{\prime}$ mass, 
as seen in Fig.~\ref{Minvetap}. 

%%%%%%%%%%%%%%%%%%% energia dei sette clusters %%%%%%%
%%%%%%%%%%%%%%%%%%%%%%%%%%

%%%%%%%%%%%%%%%%%%% massa invariante delle sette combinazioni %%%%%%%

%%%%%%%%%%%%%%%%%%%%%%%%%%%%%%%%%%%%%%%%%%%%%

\begin{figure}[htb]
\begin{center}
\begin{tabular}{cc}
\mbox{\epsfig{file=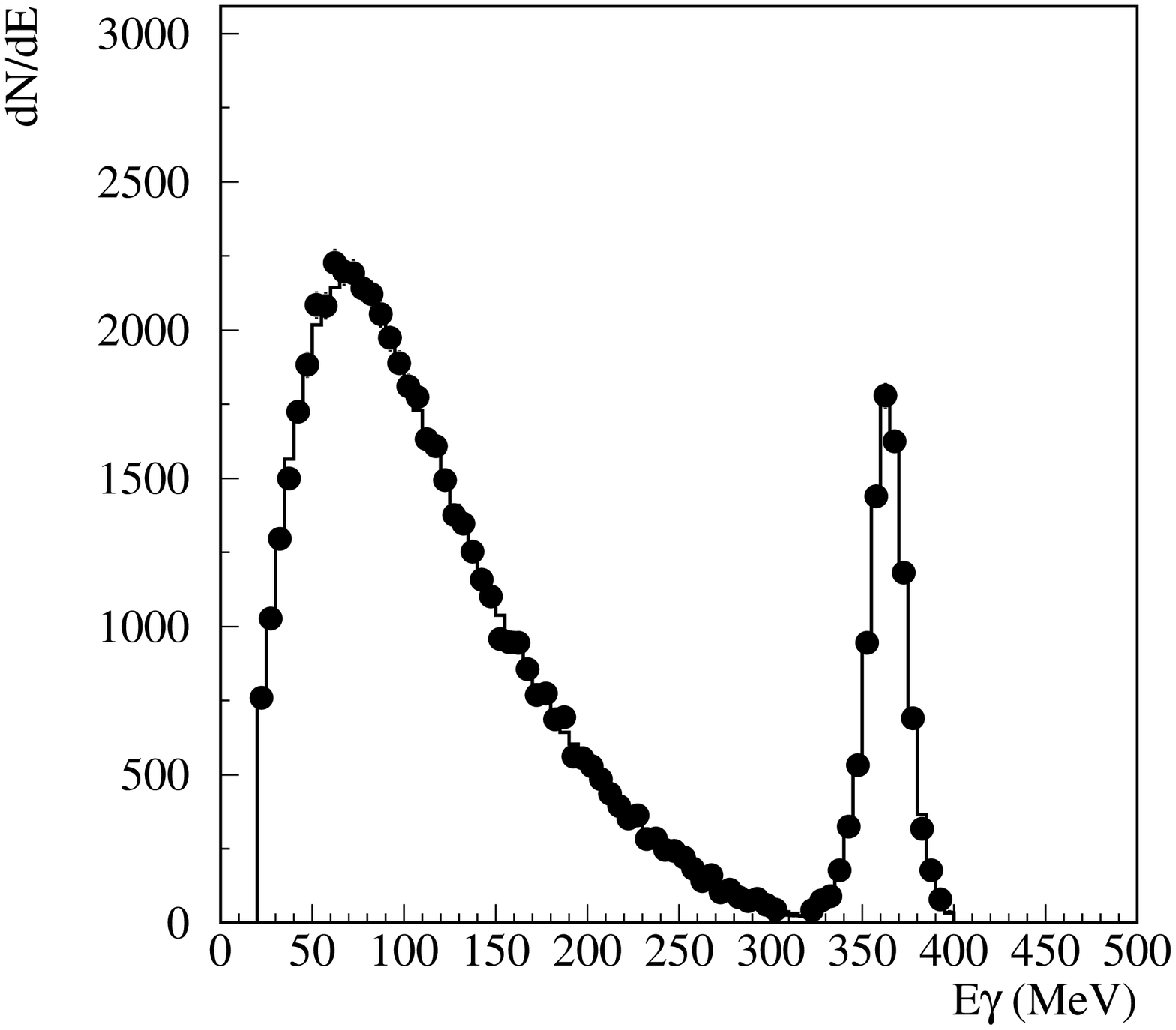,width=7.0cm}}&
\mbox{\epsfig{file=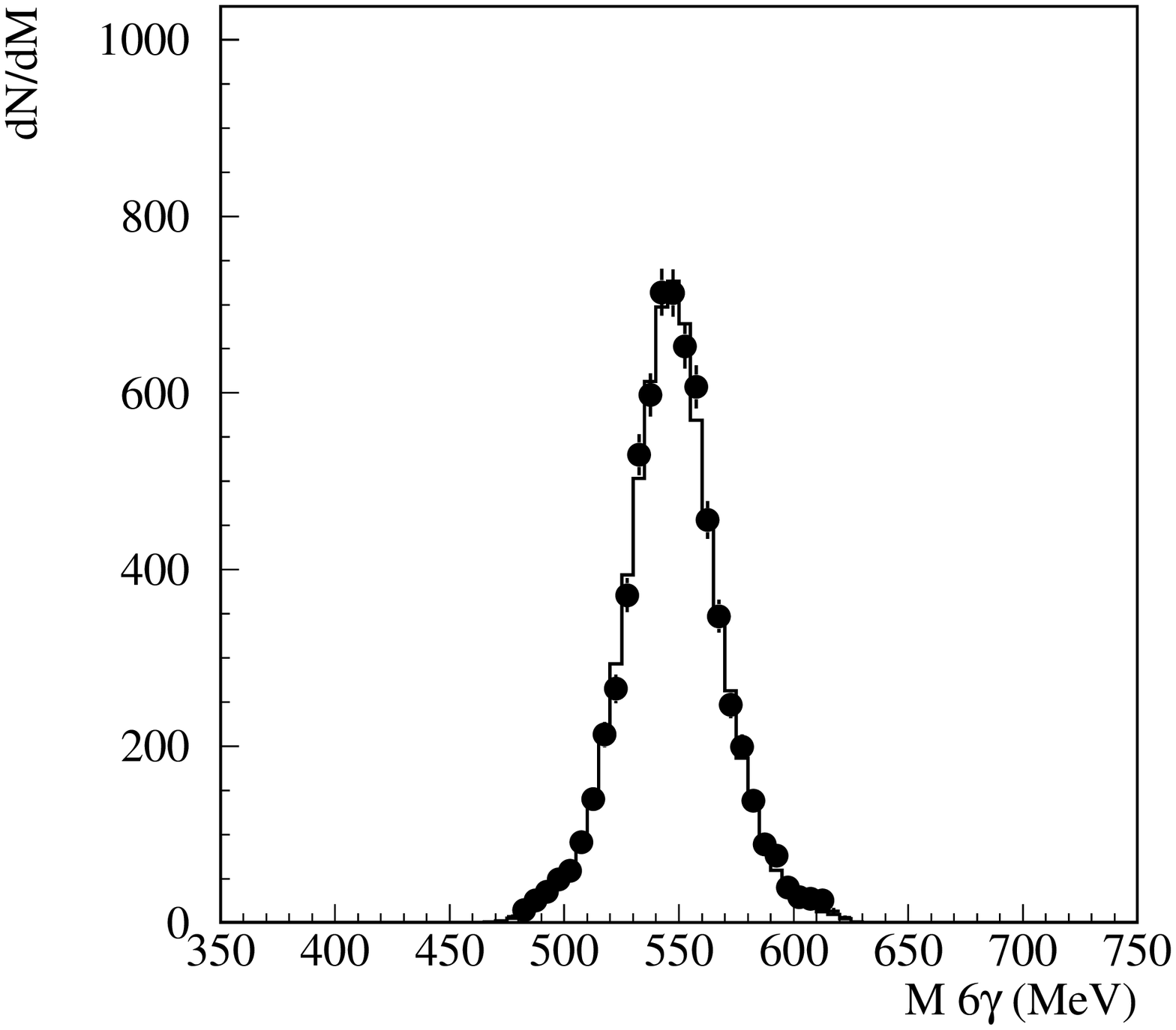,width=7.0cm}}\\
\end{tabular}
\end{center}
\caption{{\footnotesize $\phi \to \eta \gamma$. Left: Data-MC comparison of 
$E_{\gamma}$ distributions.
Right: Data-MC comparison of $M(6\gamma)$ distributions. In both panels,
histogram shows MC; dots are data.}}
\label{Minveta}
\end{figure}

In contrast to the case of $\phi \to \eta^{\prime}\gamma$ decay, 
in $\phi \to \eta \gamma$
decay the recoil photon is the most energetic photon in the event.
As a result, the recoil photons give rise to a distinct peak in the 
photon-energy spectrum.
To select process (\ref{eta}), we additionally require the identification
of the recoil photon with $320~\text{MeV}<E_{\gamma}<400~\text{MeV}$.
This cut is fully efficient for signal events and allows us to build the 
$6\gamma$ invariant mass, as seen from Fig.~\ref{Minveta}.
In all, we select $N_{\eta\gamma} = 1\,665\,000$ events.
The event selection efficiency after FLS is evaluated from MC; its value is
$\varepsilon_{\eta}=(33.66\pm0.01)\%$.
The FLS efficiency measured using data is $\epsilon_{\eta}^{FLS}=(97.88\pm0.10)\%$.
%%%%%%%%%%%%%%%%%%% eta invariant mass %%%%%%%
%%%%%%%%%%%%%%%%%%%%%%%%%%%%%%%%%%%%%%%%%%%%%

\section{Results}
\label{res}
We evaluate the ratio of the two branching fractions 
$R_{\phi}=\text {BR}(\phi \to \eta^{\prime} \gamma)/\text {BR}(\phi \to \eta \gamma)$ using the relation
\begin{equation}
R_{\phi} =
\frac{N_{\eta^{\prime}\gamma}}{N_{\eta\gamma}}
\frac{\varepsilon_{\eta}}
{\varepsilon_{\eta^{\prime}}}\cdot C_{FLS}
\cdot\frac{\text {BR}_{\eta\gamma}}
{\text {BR}_{\eta^{\prime}\gamma}}\cdot K_{\rho},
\end{equation}
where $\text {BR}_{\eta\gamma}$ is the branching fraction $\text {BR}(\eta\to\pi^0\pi^0\pi^0)$
and $\text {BR}_{\eta^{\prime}\gamma}$ is $\text {BR}(\eta^{\prime}\to\pi^+\pi^-\eta)\cdot
\text {BR}(\eta\to\pi^0\pi^0\pi^0)+\text {BR}(\eta^{\prime}\to\pi^0\pi^0\eta)\cdot
\text {BR}(\eta\to\pi^+\pi^-\pi^0)$; $\varepsilon_{\eta^{\prime}}$ and $\varepsilon_{\eta}$
are the overall selection efficiencies respectively for $\phi \to \eta^{\prime} \gamma$
and for $\phi \to \eta \gamma$; and $C_{FLS}=\epsilon_{\eta}^{FLS}/\epsilon_{\eta^{\prime}}^{FLS}$
is the ratio of the FLS efficiencies.
The factor $K_{\rho}$ accounts for interference between the amplitudes
$A(\phi\rightarrow\eta(\eta^{\prime})\gamma)$ and
$A(\rho\rightarrow\eta(\eta^{\prime})\gamma)$,
and has been evaluated in the Born approximation in the
manner of Ref.~\cite{Fabio2}.
This correction depends on $\sqrt s$, for which a mean value over run
conditions has been 
evaluated; the standard deviation is taken as a contribution to the 
systematic uncertainty. 
The final value $K_{\rho}=0.95\pm0.01$ takes into account also radiative 
corrections to the $\phi$ cross section \cite{Mario}.

The systematic uncertainty on $R_{\phi}$ is dominated by the uncertainty on the intermediate
branching ratios of the $\eta^{\prime}$. The evaluation of the systematic uncertainty on the ratio of analysis
efficiencies is obtained by studying the samples and the stability of the results
with respect to (i) the first level selection, (ii) the tracking and vertex efficiencies,
and (iii) the cuts applied to select the final sample.
The contribution to the uncertainty due to FLS comes from the uncertainty on the efficiency
measured from data using a small subsample of minimum-bias events reconstructed from raw
data without the FLS filters.
The error due to the tracking and vertex efficiencies is estimated from the data-MC discrepancy
observed for a control sample of $\phi \to \pi^+ \pi^- \pi^0$ events with charged-pion momenta
in the range covered by our analysis.
A sizeable contribution to the uncertainty due to the final selection cuts comes from the inclusiveness
of the cut on the $\chi^2$ probability from the kinematic fit, since the $\chi^2/N_{\rm dof}$
distribution has a pronounced tail.
We estimate this effect by moving the cut and evaluating the maximum variation of the ratio
$\Delta(N/\epsilon)/(N/\epsilon)$ and its effect on $R_{\phi}$.
The systematic error on $N_{\eta^{\prime}\gamma}/N_{\eta\gamma}$ comes from the background
subtraction. All the contributions are summarized in Table~\ref{tabvalue}.
\begin{table}[htb]
\vspace{+20 pt}
\caption{Contributions to the systematic error on $R_{\phi}$}
\label{tabvalue}
\begin{tabular}{@{}lcl}
\hline
Quantity with systematic source & Systematic error &Source\\
\hline\hline
$N_{\eta^{\prime}\gamma}/N_{\eta\gamma}$  & $1.3\%$ & Background   \\ \hline
$C_{FLS}$& $1.0\% $& Preselection \\\hline
 &  $1.0\%$ & Tracking efficiency\\
$\varepsilon_{\eta}/\varepsilon_{\eta^{\prime}}$ &  $1.0\%$ & Vertex efficiency\\
 &  $1.5\%$ & Prob($\chi^2$)\\\hline
$\text{BR}_{\eta\gamma}/\text{BR}_{\eta^{\prime}\gamma}$  &$2.8\%$ & \\\hline
$K_{\rho}$ & $1.0\%$&\\\hline
  \hline 
Total &  $4\%$&\\
\hline
\end{tabular}\\[2pt]
\vspace{+20 pt}
\end{table}
%%%%%%%%%%%%%%%%%%%%%%%%%%%%%%%%%%%%%%%%%%%%%%%%%%%%%%%%%%%%%%%%%%%%%%%%%%%%%%%%%%%%%%

The final result is:
\begin{equation}
R_{\phi} = \SN{(4.77 \pm 0.09_{\text{stat}} \pm 0.19_{\text{syst}})}{-3}
\end{equation}
Using the value of $\text{BR}(\phi \rightarrow \eta \gamma)=(1.301\pm0.024)\%$ 
from the PDG \cite{PDG} we find:
\begin{equation}
\text {BR}(\phi \rightarrow \eta^{\prime} \gamma) =
\SN{(6.20\pm 0.11_{\text{stat}}\pm 0.25_{\text{syst}})}{-5}
\end{equation}

The value of $R_{\phi}$ can be related to the pseudoscalar mixing angle.
In the quark-flavor basis, the $\eta$-$\eta^{\prime}$ system can be 
parameterized in terms of just one angle:
\begin{align}
 \bra{\eta} &= \cos \varphi_P\bra{q\bar{q}} + \sin \varphi_P \bra{s\bar{s}}, \nonumber \\
 \bra{\eta^{\prime}} &= -\sin \varphi_P\bra{q\bar{q}} +\cos \varphi_P \bra{s\bar{s}}. \nonumber
\end{align}
where $\bra{q\bar{q}}=(1/\sqrt{2})\bra{u\bar{u}+d\bar{d}}$.
Using the approach of Ref.~\cite{Bramon,Rafael}, where SU(3)-breaking is taken
into account via the constituent-quark mass ratio $m_s/{\bar m}$, $R_{\phi}$ is
given by the following expression:
\begin{equation}
R_{\phi} = \frac {\text{BR}(\phi \rightarrow \eta^{\prime} \gamma)} {\text{BR}(\phi \rightarrow \eta \gamma)} =  \cot^{2} \varphi_{P} \left(
1-\frac{m_s}{{\bar m}}\frac{C_{NS}}{C_{S}}\frac{\tan \varphi_V} {\sin2\varphi_{P}}\right )^2
\left( \frac{p_{\eta^{\prime}}}{p_{\eta}}  \right )^3
\label{R}
\end{equation}
where $\varphi_V=3.4^{\circ}$ is the mixing angle for vector mesons, $p_{\eta(\eta^{\prime})}$
is the ${\eta(\eta^{\prime})}$ momentum in the $\phi$ center-of-mass, and the two parameters
$C_{NS}$ and $C_{S}$ (see Table~\ref{tabpar}) represent the effect of the OZI-rule, which
reduces the vector and pseudoscalar wave-function overlap \cite{Rafael}.
From Eq.~(\ref{R}) we obtain the following result:
\begin{equation}
\varphi_{P} = (41.4\pm0.3_{\text{stat}}\pm0.7_{\text{syst}}\pm0.6_{\text{th}})^{\circ}.
\end{equation}
The theoretical uncertainty on the mixing angle has been evaluated from the maximum
variation induced by the spread of the values for $m_s/\bar{m}$,
$C_{NS}$, and $C_{S}$.
In the traditional approach, $\eta$-$\eta^{\prime}$ mixing is parameterized in the
octet-singlet basis; in this basis the value of the mixing angle becomes:
$\theta_P =\varphi_{P}-\arctan \sqrt 2=(-13.3\pm0.3_{\text{stat}}\pm0.7_{\text{syst}}
\pm0.6_{\text{th}})^{\circ}$.

In QCD, gluons may form a bound state, called gluonium, that can mix with 
neutral mesons.
While the $\eta$ is well understood as an SU(3)-flavor octet meson
with a small singlet admixture, the $\eta^{\prime}$ is a good candidate to have
a sizeable gluonium content. If we allow for non-zero $\eta^{\prime}$ 
gluonium content, we have the following parameterization:
\begin{equation*}
 \bra{\eta^{\prime}} = X_{\eta^{\prime}}\bra{q\bar{q}} + Y_{\eta^{\prime}}\bra{s\bar{s}} + Z_{\eta^{\prime}}\bra{\text{gluons}}
\end{equation*}
where $Z_{\eta^{\prime}}$ parameterizes mixing with gluonium.
Normalization implies $X_{\eta^{\prime}}^2+Y_{\eta^{\prime}}^2+Z_{\eta^{\prime}}^2=1$ with
\begin{equation}
\begin{split}
X_{\eta^{\prime}}& = \cos \phi_G \sin\varphi_P \\
Y_{\eta^{\prime}}& = \cos \phi_G \cos\varphi_P \\  
Z_{\eta^{\prime}}& = \sin \phi_G 
\end{split}
\label{parglue}
\end{equation}
where $\phi_G$ is the mixing angle for the gluonium contribution.
Non-zero $\eta^\prime$ gluonium content would imply
\begin{equation}
X_{\eta^{\prime}}^2+Y_{\eta^{\prime}}^2<1
\end{equation}
and Eq.~(\ref{R}) would have to be rewritten
\begin{equation}
R_{\phi} = \cot^{2} \varphi_{P} \cos^{2} \phi_{G} \left(
1-\frac{m_s}{{\bar m}}\frac{C_{NS}}{C_{S}}\frac{\tan \varphi_V} {\sin2\varphi_{P}}\right )^2
\left( \frac{p_{\eta^{\prime}}}{p_{\eta}}  \right )^3
\label{eqR}
\end{equation}

\begin{figure}[htb]
\begin{center}
 \mbox{\epsfig{file=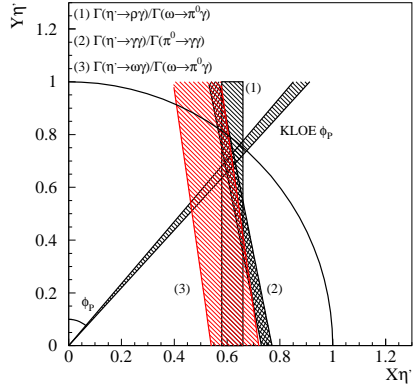,width=7.0cm}}
\end{center}
\caption{{\footnotesize Bounds on $X_{\eta^{\prime}}$, $Y_{\eta^{\prime}}$.
(1): $\Gamma(\eta^{\prime}\to \rho \gamma)/\Gamma(\omega\to \pi^0 \gamma)$;
(2): $\Gamma(\eta^{\prime}\to \gamma \gamma)/\Gamma(\pi^0\rightarrow \gamma \gamma)$; 
KLOE $\varphi_P$: $\Gamma(\phi \rightarrow \eta^{\prime}\gamma)/\Gamma(\phi \rightarrow \eta\gamma)$;
(3): $\Gamma(\eta^{\prime}\rightarrow \omega \gamma)/\Gamma(\omega\rightarrow \pi^0 \gamma)$.}}
\label{glue}
\end{figure}

We may use SU(3) relations as proposed in Refs.~\cite{Rosner,Rafael,kou} 
to further constrain
$X_{\eta^{\prime}}$ and $Y_{\eta^{\prime}}$. 
With the parameterization of Eq.~(\ref{parglue}), 
these relations may be written as:
\begin{align*}
\Gamma(\eta^{\prime}\rightarrow \gamma \gamma)/\Gamma(\pi^0\rightarrow \gamma \gamma)&=
\frac{1}{9} \left (\frac{m_{\eta^{\prime}}}{m_{\pi}} \right )^3
 (5\cos \phi_G \sin\varphi_P+\sqrt 2 \frac{f_q}{f_s} \cos \phi_G
 \cos\varphi_P)^2\\
\Gamma(\eta^{\prime}\rightarrow \rho \gamma)/\Gamma(\omega\rightarrow \pi^0 \gamma)&=\frac{C_{NS}}{\cos\varphi_V}\cdot
3 \left (\frac{m_{\eta^{\prime}}^2-m_{\rho}^2}{m_{\omega}^2-m_{\pi}^2}
\frac{m_{\omega}}{m_{\eta^{\prime}}}
\right )^3 \cos^2 \phi_G \sin^2\varphi_P\\
\Gamma(\eta^{\prime}\rightarrow \omega \gamma)/\Gamma(\omega\rightarrow \pi^0 \gamma)&=\frac{1}{3}\left (\frac{m_{\eta^{\prime}}^2-m_{\omega}^2}{m_{\omega}^2-m_{\pi}^2}\frac{m_{\omega}}{m_{\eta^{\prime}}}\right )^3 
 [C_{NS}\cdot \cos \phi_G \sin\varphi_P\\
 &+2\frac{m_s}{{\bar m}}C_{S}\cdot \tan\varphi_V\cdot\cos\phi_G\cos\varphi_P]^2
\end{align*}
Figure~\ref{glue} shows the above constraints graphically in the
$(X_{\eta^{\prime}}, Y_{\eta^{\prime}})$ plane.
The constraint from the KLOE measurement of $R_{\phi}$ is shown in the
hypothesis of no $\eta^\prime$ gluonium content.
To allow for gluonium, we have minimized a $\chi^2$ function with 
$\cos^2 \phi_G$ and $\cos^2 \varphi_P$ as free parameters by imposing 
the above constraints and including in the error matrix
the uncertainties on the other parameters as shown in Table~\ref{tabpar}.
The values for $f_q$ and $f_s$ are from Ref.~\cite{Feld}; all other values 
are from \cite{Rafael}. We assume no correlations between the parameters.

%%%%%%%%%%%%%%%%%%%% colla1 %%%%%%%
%%%%%%%%%%%%%%%%%%%%%%%%%%%%%%%%%%%%%%%%%%%%

%%%%%%%%%%%%%%%%%%%%%%%%%%%%%%Table 2.3%%%%%%%%%%%%%%%%%%%%
\begin{table}[htb]
\vspace{+20 pt}
\caption{Fixed parameters used in the fit for $\cos^2\varphi_G$ and 
$\cos^2\varphi_P$} 
\label{tabpar}
\begin{tabular}{@{}lccccc}
\hline
Parameter& $f_q$    &$f_s$    &$C_{NS}$ &$C_{S}$  &$\frac{m_s}{{\bar m}}$ \\ \hline
Value & $1\pm0.01$ &  $1.4\pm0.014$ &  $0.91\pm0.05$ &  $0.89\pm0.07$ & $1.24\pm0.07$\\    \hline 
\end{tabular}\\[2pt]
\vspace{+20 pt}
\end{table}
%%%%%%%%%%%%%%%%%%%%%%%%%%%%%%%%%%%%%%%%%%%%

The solution in the hypothesis of no gluonium content, i.e. $\cos^2\phi_{G}=1$,
yields $\varphi_{P} = (41.5^{+0.6}_{-0.7})^{\circ}$.
The value of $\chi^2/N_{\rm dof}$ is $11.34/3$, corresponding to a 
probability of 0.01, and suggests a possible non-zero value for 
$Z^2_{\eta^{\prime}}$.

\begin{figure}[htb]
\begin{center}
\mbox{\epsfig{file=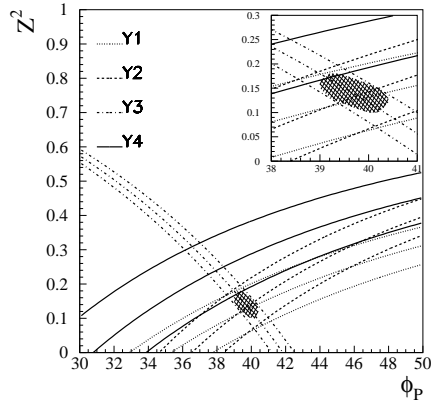,width=7cm}} 
\end{center}
\caption{{\footnotesize The four constraints in the $(Z^2, \varphi_{P})$ plane:
$y_1=\Gamma(\eta^{\prime}\rightarrow \gamma \gamma)/\Gamma(\pi^0\rightarrow \gamma \gamma)$,
$y_2=\Gamma(\eta^{\prime}\rightarrow \rho \gamma)/\Gamma(\omega\rightarrow \pi^0 \gamma)$,
$y_3=R_{\phi}$, and $y_4=\Gamma(\eta^{\prime}\rightarrow \omega \gamma)/
\Gamma(\omega\rightarrow \pi^0 \gamma)$.
The allowed region for the solution with gluonium is shown.}}
\label{glue1}
\end{figure}

The solution allowing for gluonium is $\cos^2\phi_{G}=0.86\pm0.04$
and $\cos^2\varphi_{P}=0.592\pm0.012$, from which 
$\varphi_{P} = (39.7\pm0.7)^{\circ}$
and $Z^2_{\eta^{\prime}} = 0.14\pm0.04$, 
which means $|\varphi_{G}| = (22\pm3)^{\circ}$.
In this case, $\chi^2/N_{\rm dof} =1.42/2$, for a probability of 0.49.
When the constraint from $\eta^{\prime}\rightarrow \omega \gamma$ decay
is relaxed, similar results are obtained:
$\varphi_{P} = (39.8\pm0.8)^{\circ}$ and
$Z^2_{\eta^{\prime}} = 0.13\pm0.04$, 
while the $\chi^2$ probability improves to 0.97.
Figure~\ref{glue1} shows the four constraints in the 
$(Z^2_{\eta^{\prime}}, \varphi_{P})$
plane, together with the allowed region for the solution with gluonium.

In the octet-singlet basis, the value of the $\eta$-$\eta^{\prime}$ mixing
angle $\varphi_{P} = (39.7\pm0.7)^{\circ}$ becomes 
$\theta_P = (-15.0\pm0.7)^{\circ}$,
which is in agreement with the value $\theta_P = (-15.9\pm1.2)^{\circ}$ 
measured by BES-II \cite{BES} using the branching fractions 
$BR(J/\psi \to \eta^{\prime} \gamma)$ and
$BR(J/\psi \to \eta \gamma)$.
These results for $\varphi_{P}$ are consistent with those coming from 
other processes as reviewed in Ref.~\cite{Feld},
which quotes an average value $\varphi_{P} = (39.3\pm1.0)^{\circ}$.

\section{Conclusion}
\label{conc}
Analyzing about $\SN{1.4}{9}$ $\phi$ mesons collected by KLOE at the
\DAFNE\ collider, we have obtained a precise new measurement of the ratio
$R_{\phi}=\text{BR}(\phi \to \eta^{\prime} \gamma)/\text{BR}(\phi \to \eta \gamma)=
\SN{(4.77\pm0.09_{\text{stat}}\pm0.19_{\text{syst}})}{-3}$.
From this measurement, we obtain the most precise value of the 
pseudoscalar mixing angle in the flavor basis: 
$\varphi_P=(41.4\pm0.3_{\text{stat}}\pm0.7_{\text{syst}}
\pm0.6_{\text{th}})^{\circ}$.
Allowing for gluonium in the $\eta^{\prime}$, a fit to our result together
with other measurements yields $\varphi_{P} = (39.7\pm0.7)^{\circ}$
and a $\eta^{\prime}$ gluonium content of $Z^2_{\eta^{\prime}} = 0.14\pm0.04$.

\section*{Acknowledgements}
We thank the \DAFNE\ team for their efforts in maintaining 
low background running 
conditions and their collaboration during all data taking. 
We want to thank our technical staff: 
G.F.~Fortugno for his dedicated work to ensure an efficient operation of 
the KLOE Computing Center; 
M.~Anelli for his continuous support of the gas system and the safety of
the detector; 
A.~Balla, M.~Gatta, G.~Corradi and G.~Papalino for the maintenance of the
electronics;
M.~Santoni, G.~Paoluzzi and R.~Rosellini for the general support of the
detector; 
C.~Piscitelli for his help during major maintenance periods.
This work was supported in part by DOE grant DE-FG-02-97ER41027; 
by EURODAPHNE, contract FMRX-CT98-0169; 
by the German Federal Ministry of Education and Research (BMBF) contract 06-KA-957; 
by Graduiertenkolleg `H.E. Phys. and Part. Astrophys.' of Deutsche Forschungsgemeinschaft,
Contract No. GK 742; 
by INTAS, contracts 96-624, 99-37; 
and by the EU Integrated Infrastructure
Initiative HadronPhysics Project under contract number
RII3-CT-2004-506078.

% Bibliographic references with the natbib package:
% Parenthetical: \citep{Bai92} produces (Bailyn 1992).
% Textual: \citet{Bai95} produces Bailyn et al. (1995).
% An affix and part of a reference:
%   \citep[e.g.][Ch. 2]{Bar76}
%   produces (e.g. Barnes et al. 1976, Ch. 2).

\end{document}